\documentclass[reprint,superscriptaddress,amsmath,amssymb,aip]{revtex4-1}
\pdfoutput=1

\usepackage{graphicx}
\usepackage{gensymb}
\usepackage{hyperref}
\usepackage{float}
\usepackage{outlines}
\usepackage{enumitem}
\usepackage{upgreek}

\usepackage{color,soul} 

\begin{document}

\title{Superconducting nanowire single-photon detector with integrated impedance-matching taper}

\author{Di Zhu}
\author{Marco Colangelo}
\affiliation{Research Laboratory of Electronics, Massachusetts Institute of Technology, Cambridge, MA, United States}

\author{Boris A. Korzh}
\affiliation{Jet Propulsion Laboratory, California Institute of Technology, Pasadena, CA, United States}

\author{Qing-Yuan Zhao}\affiliation{Research Laboratory of Electronics, Massachusetts Institute of Technology, Cambridge, MA, United States}

\author{Simone Frasca}
\affiliation{Jet Propulsion Laboratory, California Institute of Technology, Pasadena, CA, United States}

\author{Andrew E. Dane}\affiliation{Research Laboratory of Electronics, Massachusetts Institute of Technology, Cambridge, MA, United States}

\author{Angel E. Velasco}
\author{Andrew D. Beyer}
\author{Jason P. Allmaras}
\author{Edward Ramirez}
\affiliation{Jet Propulsion Laboratory, California Institute of Technology, Pasadena, CA, United States}

\author{William J. Strickland}
\author{Daniel F. Santavicca}\affiliation{University of North Florida, Jacksonville, FL, United States}

\author{Matthew D. Shaw}\affiliation{Jet Propulsion Laboratory, California Institute of Technology, Pasadena, CA, United States}

\author{Karl K. Berggren}\email{berggren@mit.edu}
\affiliation{Research Laboratory of Electronics, Massachusetts Institute of Technology, Cambridge, MA, United States}

\date{22 October 2018}
\begin{abstract} 
	Conventional readout of a superconducting nanowire single-photon detector (SNSPD) sets an upper bound on the output voltage to be the product of the bias current and the load impedance, $I_\mathrm{B}\times Z_\mathrm{load}$, where $Z_\mathrm{load}$ is limited to $50\,\Omega$ in standard r.f. electronics. Here, we break this limit by interfacing the 50\,$\Omega$ load and the SNSPD using an integrated superconducting transmission line taper. The taper is a transformer that effectively loads the SNSPD with high impedance without latching. It increases the amplitude of the detector output while preserving the fast rising edge. Using a taper with a starting width of 500\,nm, we experimentally observed a 3.6$\times$ higher pulse amplitude, 3.7$\times$ faster slew rate, and 25.1\,ps smaller timing jitter. The results match our numerical simulation, which incorporates both the hotspot dynamics in the SNSPD and the distributed nature in the transmission line taper. The taper studied here may become a useful tool to interface high-impedance superconducting nanowire devices to conventional low-impedance circuits. 
\end{abstract}


\maketitle

The superconducting nanowire single-photon detector (SNSPD) is the leading single-photon detection technology at infrared wavelengths.\cite{Hadfield2009,Natarajan2012}  With exceptional performance,\cite{Marsili2013, Korzh2018, Wollman2017, Vetter2016, Marsili2012} it has played an essential role in various applications, especially quantum information science\cite{Yin2016, Qiang2018} and deep-space optical communication.\cite{Grein2015} 

A common problem with SNSPDs is their low output voltage and signal-to-noise ratio (SNR), which has been a limiting factor in detector timing jitter.\cite{Wu2017} A simple lumped-circuit model dictates that the output voltage from the nanowire cannot exceed $I_\mathrm{B}\times Z_\mathrm{load}$, where $I_\mathrm{B}$ is the bias current and $Z_\mathrm{load}$ is the load impedance.\cite{Kerman2009} $I_\mathrm{B}$ is limited by the nanowire's switching current at the $\upmu$A range. $Z_\mathrm{load}$ is set by the input impedance of the coaxial cable and r.f. electronics, which is conventionally 50\,$\Omega$. To improve readout SNR, significant progress has been made in developing cryogenic amplifiers with low noise, dissipation and cost, e.g. using silicon germanium and gallium arsenide transistors.\cite{Bardin2014, Mani2014, Wan2017, Cahall2018} Digital readout circuits built directly from superconducting electronics, such as nanocryotrons~\cite{McCaughan2014} and single flux quantum (SFQ) circuits,\cite{Terai2010, Ortlepp2011} have also been demonstrated. These integrated superconducting circuits are low-noise and scalable, but usually require additional biasing and suffer from leakage current and crosstalk.

An alternative approach to increase the output signal is to increase $Z_\mathrm{load}$. Compared to a standard 50\,$\Omega$ load, a high-impedance load is often more desirable---it not only increases the detector output, but also enables direct mapping of hotspot resistance and photon number/energy resolution.\cite{Bell2007, Kitaygorsky2009, Jahanmirinejad2012} However, high-impedance loading is difficult to achieve in practice. The lack of high-impedance coaxial cables makes it necessary to place the high-impedance amplifiers close to the detectors (at the low-temperature stage), which imposes a more stringent power budget. More importantly, even if a high-impedance amplifier is available,\cite{Wan2017} loading a standard SNSPD directly with high impedance can lead to latching.\cite{Kerman2009} 

In this work, without the need of high-impedance cryogenic amplifiers or any active circuit elements, we break the $I_\mathrm{B}\times 50\,\Omega$ limit by using an integrated superconducting transmission line taper. The taper gradually transforms its characteristic impedance from k$\Omega$ to 50\,$\Omega$, which effectively loads the SNSPD with a k$\Omega$ impedance without latching. We designed the taper to be a co-planar waveguide (CPW) and fabricated it from the same superconducting thin film as the SNSPD. Using a taper with a starting width of 500\,nm and nominal passband from 200\,MHz, we experimentally observed 3.6$\times$ higher output voltage and no added noise compared to the non-tapered reference device. This voltage gain is equivalent to a 11\,dB passive, dissipation-free cryogenic amplifier. Despite its large inductance, the taper preserves the detector's fast rising edge, resulting in an increased slew rate and reduced timing jitter (from 48.9 ps to 23.8 ps). The integrated impedance taper demonstrated here is useful for interfacing high-impedance nanowire-based devices to conventional low-impedance components, such as memory, and electrical or optical modulators. 

\begin{figure}
	\includegraphics[width = 3.2 in]{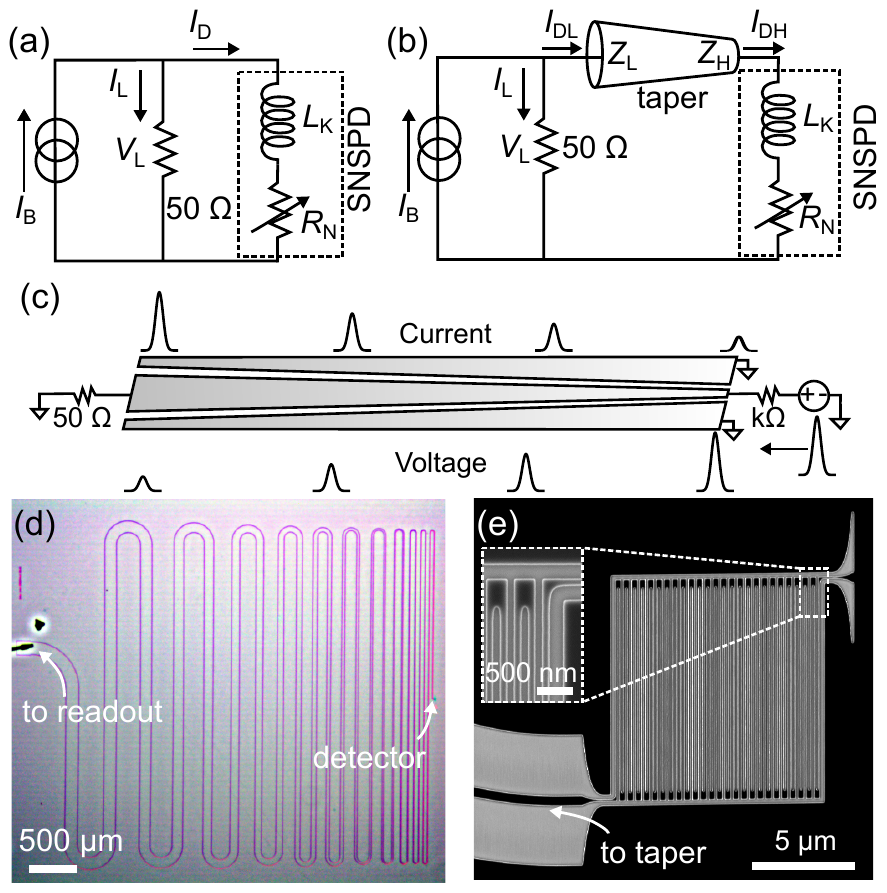}
	\caption{Circuit diagram and micrographs of the tapered SNSPD readout. (a) A circuit diagram of a conventional SNSPD readout. (b) A circuit diagram of a tapered readout. The taper loads the SNSPD with high impedance while interfacing at its other end to the readout electronics at 50\,$\Omega$, resulting in a larger output voltage. (c) Schematic diagram of a co-planar waveguide transmission line taper. When an electrical pulse is launched from the high-impedance end, its voltage drops but current increases while traveling towards the low-impedance end. (d) An optical micrograph of the integrated transmission line taper. Light area: NbN; red outlines: substrate. (e) A scanning electron micrograph of the SNSPD. Dark area: NbN; light area: substrate.}
	\label{fig:fig1}
\end{figure}

Figure~\ref{fig:fig1}(a) shows a circuit model of a conventional SNSPD readout circuit, where the detector is modeled as a kinetic inductor $L_\mathrm{K}$ in series with a time-dependent variable resistor $R_\mathrm{N}$. When an incident photon triggers the detector, $R_\mathrm{N}$ switches from 0 to $\approx$k$\Omega$ within $\approx$100s of ps and diverts the bias current to the load. The evolution of $R_\mathrm{N}$ is determined by the non-linear electrothermal feedback in the detector.\cite{Yang2007,Kerman2009} The currents from the bias source ($I_\mathrm{B}$), in the nanowire ($I_\mathrm{D}$), and to the load ($I_\mathrm{L}$) simply follow Kirchhoff's law $I_\mathrm{L}=I_\mathrm{B}-I_\mathrm{D}$. The maximum $I_\mathrm{L}$ is therefore limited to $I_\mathrm{B}$, corresponding to the case where $R_\mathrm{N}$ pushes all of the current out of the nanowire ($I_\mathrm{D}=0$). The output voltage on the load thus can not exceed $I_\mathrm{B}\times 50\,\Omega$. In practice, due to the electro-thermal feedback,\cite{Kerman2009} $I_\mathrm{D}$ usually has some remainder, which depends on the bias current, kinetic inductance, and thermal constants of the materials. 

Figure~\ref{fig:fig1}(b) shows a simplified circuit diagram for the tapered readout. The taper is inserted between the SNSPD and load, with a low impedance $Z_\mathrm{L}=50\,\Omega$ on the load end and a high impedance $Z_\mathrm{H}$ on the detector end. In our implementation, the taper consists of a continuous nanowire transmission line without any dissipative elements. The taper is high-pass---it works as a transformer at high frequency but acts as an inductor at low frequency. When an incident photon triggers the SNSPD, $R_\mathrm{N}$ switches on and pushes the current away from the nanowire. Instead of diverting the current directly to the 50\,$\Omega$ load as in the conventional readout, the SNSPD injects current to the taper at $Z_\mathrm{H}$. As the electrical pulse travels towards the low impedance end, its current amplitude increases while the voltage amplitude drops, with a ratio that satisfies the change of impedance (Fig.~\ref{fig:fig1}(c)). Assuming an ideal broadband transformer with perfect impedance matching and power transmission, the current leaving the low-impedance end (to the load) $\Delta I_\mathrm{DL}$ is related to the current injected to the high-impedance end $\Delta I_\mathrm{DH}$ by $\Delta I_\mathrm{DL}^2 Z_\mathrm{L}=\Delta I_\mathrm{DH}^2 Z_\mathrm{H}$. In our transmission line taper, this relation is valid only at high frequency (passband of the taper), which dominates the rising edge of the detector pulse. In the extreme case where the SNSPD pushes all the bias current out, i.e., $\Delta I_\mathrm{DH}=I_\mathrm{B}$, the current diverted to the load can be as large as $I_\mathrm{L}=\Delta I_\mathrm{DL}=I_\mathrm{B}\sqrt{\frac{Z_\mathrm{H}}{Z_\mathrm{L}}}$, corresponding to an output voltage of $V_\mathrm{L}^\mathrm{taper}=I_\mathrm{B}\times 50\,\Omega\sqrt{\frac{Z_\mathrm{H}}{Z_\mathrm{L}}}$ and an effective voltage gain of $\sqrt{\frac{Z_\mathrm{H}}{Z_\mathrm{L}}}$ with respect to the conventional readout. In practice, when terminated with the high-impedance taper, the SNSPD latches, leaving the residual current at the hotspot current $I_\mathrm{ss}$, then resets through reflection from the taper. As we will show later, the actual voltage gain is always less than $\sqrt{\frac{Z_\mathrm{H}}{Z_\mathrm{L}}}$ due to the electro-thermal feedback and limited taper bandwidth. 

\begin{figure}
	\centering
	\includegraphics[width = 2.6 in]{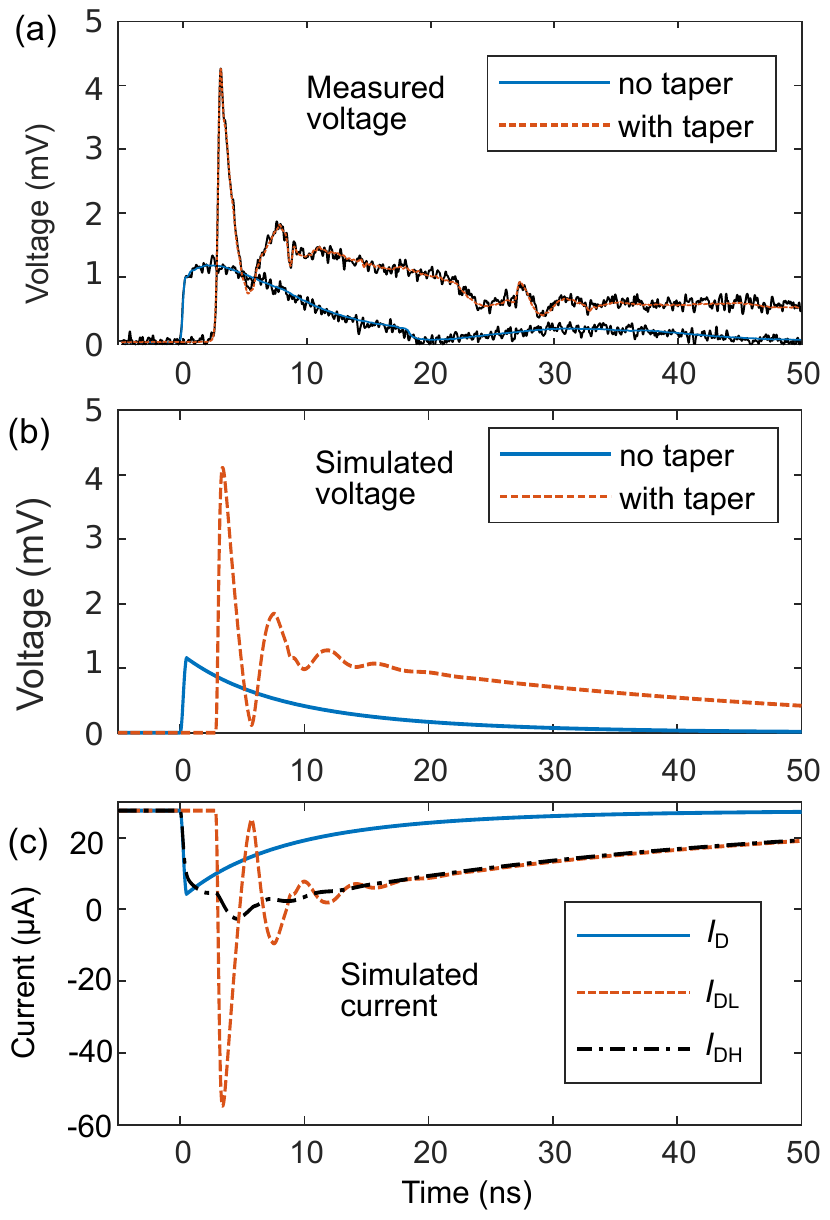}
	\caption{Measured detector pulses and comparison with SPICE simulation. (a) Measured voltage with amplifier gain removed. The black traces are single-shot waveforms, and the superimposed colored lines are averaged waveforms. (b) Simulated output voltages for both the tapered and non-tapered detectors. (c) Simulated current distributions. $I_\mathrm{D}$ is the current in the SNSPD for a non-tapered readout. $I_\mathrm{DL}$ and $I_\mathrm{DH}$ are the currents on the low-impedance and high-impedance ends of the tapers in the tapered detector, respectively.} 
	\label{fig:fig2}
\end{figure}

Figure~\ref{fig:fig2}(d) shows an optical micrograph of a fabricated SNSPD with a meandered transmission-line taper. The bright area is NbN, and the red area is the substrate, where the NbN was etched away. The NbN was sputtered at room temperature on a silicon substrate with a 300\,nm thick thermal oxide layer.\cite{Dane2017a} The film had a critical temperature of 8.1\,K and room-temperature sheet resistance of 342\,$\Omega$/sq. The sheet inductance was estimated to be 80\,pH/sq by fitting the falling edge of the output pulse from a reference detector. The nanowire fabrication process is described in Ref.~\cite{Zhao2017,Korzh2018}. The taper was made from a CPW with a fixed gap size of 3\,$\upmu$m, and a varying center conductor width from 135\,$\upmu$m (50\,$\Omega$) to 500\,nm (1.7\,k$\Omega$). Its left/wide end is wire bonded to an external circuit board, and the right/narrow end connects to the SNSPD through a 1 \,$\upmu$m-long hyperbolic taper. The SNSPD was 100\,nm wide, densely packed with a 50\% fill factor, and spanned a rectangular area of 11\,$\upmu$m$\times$10\,$\upmu$m (see Fig.~\ref{fig:fig1}(e)). A 200\,nm gap surrounded the detector region to reduce the proximity effect in fabrication. On the same chip, we also fabricated non-tapered detectors as references.

The taper was designed to be a 5672-section cascaded transformer with a lower cut-off frequency of 200\,MHz and a total electrical length of 851 mm, following the Klopfenstein profile.\cite{Klopfenstein1956} The physical length was 77.9\,mm due to the slow phase velocity of the superconducting transmission line, and the total inductance was 1.410 $\upmu$H (see SI for the simulated S paramters). This length was chosen so that the maximum reflection in the passband ($\ge200$\,MHz) would not exceed -20\,dB. The total electrical length is calculated as $l_\mathrm{e}= A c/2\pi f_\mathrm{co}$, where $f_\mathrm{co}$ is the nominal cut-off frequency and $c$ is the speed of light in vacuum.  $A$ is a factor that determines the maximum reflection in the passband and is calculated as $\cosh(A) = \rho_0/\rho_\mathrm{pb}$, where $\rho_0$ is the initial reflection coefficient (i.e., without taper) and $\rho_\mathrm{pb}$ is the maximum passband reflection coefficient (taken as 0.1 here).  For design convenience, we followed Klopfenstein's original approach and took $\rho_0=0.5\ln(Z_\mathrm{H}/Z_\mathrm{L})$ instead of $(Z_\mathrm{H}-Z_\mathrm{L})/(Z_\mathrm{H}+Z_\mathrm{L})$.\cite{Klopfenstein1956} 

The detectors were measured at 1.3\,K in a closed-cycle cryostat. Both the bias circuit and readout electronics were at room temperature. The output signal of the detectors were amplified using a 2.5\,GHz, 25 dB gain low-noise amplifier (RF BAY LNA-2500), and a 3\,dB attenuator was inserted before the amplifier to reduce reflection and prevent latching. The output pulses from the amplifier were then acquired by a 6\,GHz real-time oscilloscope (Lecroy 760Zi). The detector chip was flood illuminated using attenuated sub-ps pulsed lasers at 1550\,nm (FPL-02CCF) through an optical fiber (SMF-28e). The laser pulses were split into two arms, one to a variable attenuator then to the cryostat, and the other to a fast photodiode (Thorlabs DET08CFC) as timing references. Since the distance between the non-tapered detector and tapered detector ($\approx$ 5\,mm) was much less than the distance between the detector chip and fiber tip ($\approx$ 10\,cm), we expect the difference in photon arrival time to be $<1$\,ps. Both the tapered and non-tapered detectors had a switching current of 30\,$\upmu$A, and were biased at 27.5\,$\upmu$A throughout the measurement. 

Figure~\ref{fig:fig2}(a) shows the measured pulse shapes from the reference and tapered detectors. The amplifier gain was removed to better compare with simulations. To avoid phase distortion in reconstructing the unamplified pulses, we used a weighted gain, $\bar{G}=\int \mathrm{d}f \mathit{PSD}(f)G(f)/\int\mathrm{d}f\mathit{PSD}(f)$, where $\mathit{PSD}(f)$ is the power spectral density of the pulse, and $G(f)$ is the measured system gain spectrum (see SI for details). $\bar{G}$ was calculated to be 20.5\,dB. As shown in Fig.~\ref{fig:fig2}(a), we observed a voltage gain of 3.6 and an extra delay of 2.8\,ns from the tapered device compared to the reference device (by aligning the electrical pulses to the optical references). This voltage enhancement is equivalent to a passive, excessive-noise-free 11\,dB amplifier.

We simulated the tapered detector using a SPICE model that incorporates both the hotspot dynamics in the SNSPD and the distributed nature of the transmission line taper.\cite{Berggren2018,Zhao2018a} The simulation was implemented in LTspice, a free electrical circuit modeling software. The SPICE model for SNSPD was developed by Berggren et al.\cite{Berggren2018} based on the phenomenological hostspot velocity model by Kerman et al.\cite{Kerman2009} The taper was simulated as cascaded lossless transmission lines (down-sampled to 300 sections),\cite{Zhao2018a} and each section was implemented using the LTRA model in LTspice with different length, inductance, and capacitance settings.

Figure~\ref{fig:fig2}(b) plots the simulated load voltages, showing a voltage gain of 3.5 and a delay of 2.8\,ns, as compared to the measured gain of 3.6 and delay of 2.8\,ns. The subsequent peaks in the output voltage are spaced by $\approx$4.2\,ns, which should correspond to the round trip time in the taper. The single-trip delay of the taper calculated from the reflection peaks (2.1\,ns) is shorter than the delay between the tapered detector and reference detector (2.82\,ns), because the hotspot grows for a longer period of time and to a larger resistance in the tapered detector, as can be seen in the simulated currents.

Figure~\ref{fig:fig2}(c) shows the simulated currents. For the tapered detector, the current in the nanowire ($I_\mathrm{DH}$) first drops at a similar rate as the non-tapered case ($I_\mathrm{D}$), then enters a intermediate plateau due to latching.\cite{Kerman2009} A similar current plateau and latching behavior are often observed when loading an SNSPD with a k$\Omega$ resistor. However, at $\approx$ 4.6\,ns, $I_\mathrm{DL}$ drops again and kicks the detector out of the latching state. The drop in current is from the reflection in the transmission line taper. Alternatively, it can be interpreted as the distributed capacitors in the transmission line drawing current from the SNSPD. After a few oscillations (high frequency), the current in the detector recovers with an $\tau=L/R$ exponential time constant (low frequency). Here, $L$ is the total inductance of the SNSPD and the taper (at low frequency, the taper behaves as an inductor), and $R$ is 50\,$\Omega$. The simulated currents at the high- and low-impedance ends of the taper follow our intuitive understanding on how a transformer works. In this particular detector and taper design, the  maximum counting rate (estimated as 1/3$\tau$) decreases from 40.3\,MHz ($L=414$\,nH) for the non-tapered detector to 9.1\,MHz ($L=1.824$\,$\upmu$H) for the tapered detector. We would like to point out that the added inductance, and consequently slower maximum counting rate, are currently the major drawbacks of the tapered readout. A study of the trade-off between the gain factor, taper impedance, bandwidth, and inductance can be found in the SI.

\begin{figure}
	\centering
	\includegraphics[width = 3in]{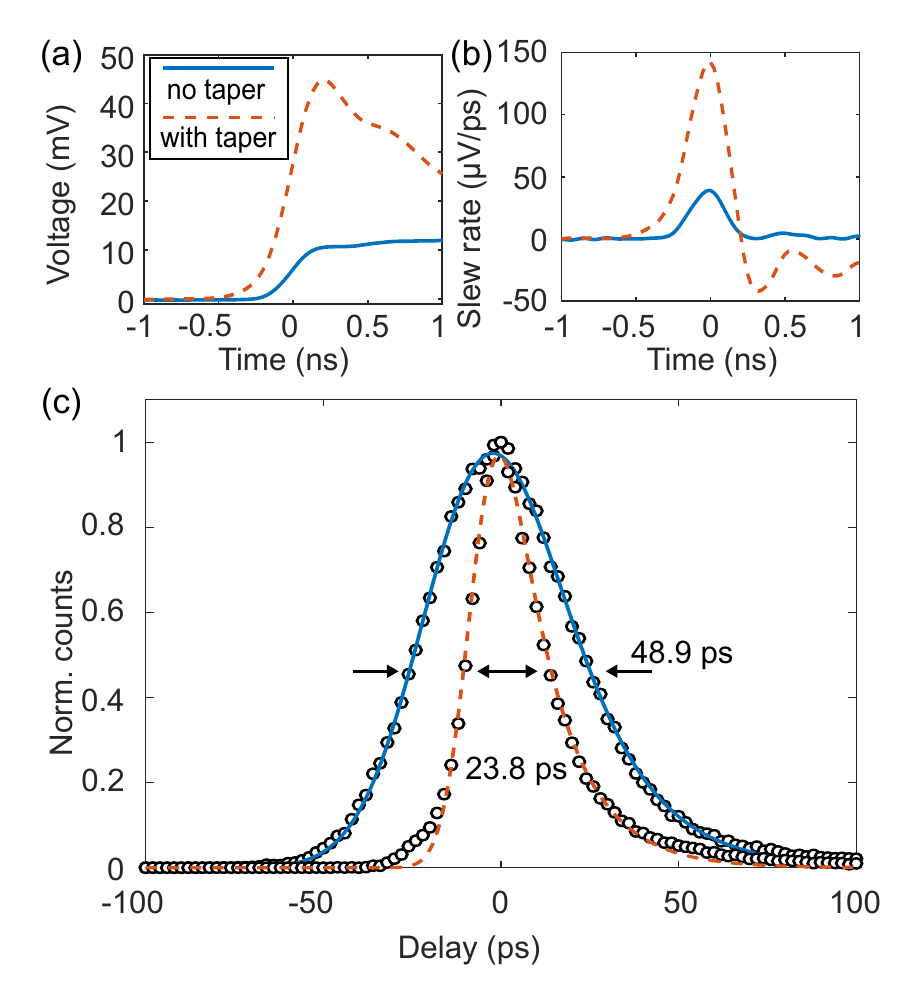}
	\caption{Experimental observation on the reduction of timing jitter as a result of faster slew rate. (a) Averaged rising edges of the detector pulses from the tapered and non-tapered detectors (amplifier gain not removed); (b) corresponding slew rate calculated as $\mathrm{d}V/\mathrm{d}t$. (c) The measured FWHM timing jitter reduced from 48.9\,ps to 23.8\,ps with the tapered readout at 1550\,nm.}
	\label{fig:fig_jitter}
\end{figure}

The impedance taper amplifies the output pulse without sacrificing the fast rising edge, resulting in a faster slew rate. Figure~\ref{fig:fig_jitter}(a) compares the averaged rising edges of the detector pulses from the reference and tapered detectors (with amplifier gain). The sampling rate was 40 GS/s. As shown in Fig.~\ref{fig:fig_jitter}(b), the maximum slew rates ($\mathrm{d}V/\mathrm{d}t$) were 39\,$\upmu$V/ps for the reference detector, but 143\,$\upmu$V/ps (3.7 times faster) for the tapered detector. 

The slew rate directly impacts the electrical noise contribution on the timing jitter, usually referred as noise jitter, $\sigma_\mathrm{noise}$.~\cite{Zhao2011,Wu2017}  We sampled the background electrical noise on the oscilloscope for both detectors by measuring the voltage at 400\,ps before the rising edge of the pulses. The noise followed a Gaussian distribution and had a standard deviation of 559\,$\upmu$V and 547\,$\upmu$V for the reference and tapered detector, respectively (see SI for details). Taking their respective fastest slew rates, we calculated that the reference detector would have a standard deviation $\sigma_\mathrm{noise}$  of 14.3\,ps, and the tapered detector would have a  $\sigma_\mathrm{noise}$ of 3.8\,ps.

We measured the jitter of the detectors following the procedure described in Ref.~\cite{Najafi2012}. The discrimination levels for time tagging were set to voltages with the fastest slew rates. Figure~\ref{fig:fig_jitter}(c) shows the instrument response function (IRF) of the reference and tapered detectors at 1550\,nm illumination wavelength. With the impedance taper, the full-width half-maximum (FWHM) jitter reduced from 48.9\,ps to 23.8\,ps. We fitted the IRF using an exponentially modified Gaussian distribution,\cite{Korzh2018} and found $\sigma = 16.8$\,ps and $1/\lambda=17.4$\,ps for the reference detector, and $\sigma = 6.5$\,ps and $1/\lambda=13.6$\,ps for the tapered detector. Here, $\sigma$ is the standard deviation of the normal distribution, $\lambda$ is the exponential decay rate. The detectors showed similar jitter reduction at 1064\,nm, where both detectors operated on the saturation plateau (see SI for the photon count rate vs. bias current curves). The FWHM jitter reduced from 47.0\,ps ($\sigma = 16.4$\,ps, $1/\lambda=15.9$\,ps) to 22.4\,ps ($\sigma = 6.2$\,ps, $1/\lambda=12.5$\,ps). We observed a leading edge tail in IRF for the tapered detector. It is likely due to the counting events from the taper or the transition region between the taper and the detector. This effect could be reduced if the optical mode is focused in the center of the detector active area, through self-aligned fiber coupling or focusing lenses.\cite{Miller2011,Bellei2016}

As a final remark, we have treated the SNSPD as a lumped element in this paper, because the nanowire was closely meandered and had a dispersion similar to an ideal inductor at the frequency of interest.\cite{Santavicca2016} Despite this choice, multi-photon absorption would generate a different hotspot resistance than the single-photon events.\cite{Bell2007, Cahall2017} The impedance taper provides an effective k$\Omega$ load, and may thus allow direct discrimination of hotspot resistance and hence photon numbers. In another scheme, where the nanowire is sparse or integrated into a transmission line,\cite{Zhao2018a} the taper can serve as an impedance-matched readout and has been used to resolve photon location and photon numbers.\cite{Zhao2017, Zhu2018} We expect the integrated taper to become a widely used tool for matching high-impedance nanowire-based devices to low-impedance systems.

Part of this work was performed at the Jet Propulsion Laboratory, California Institute of Technology, under contract with the National Aeronautics and Space Administration. Support for this work was provided in part by the JPL Strategic University Research Partnerships program, DARPA Defense Sciences Office through the DETECT program, and National Science Foundation grants under contract No. ECCS-1509486 (MIT) and No. ECCS-1509253 (UNF). D. Z. is supported by the National Science Scholarship from A*STAR, Singapore. 

%

\widetext
\clearpage
\begin{center}
	\textbf{\large Supplementary Information}
\end{center}

\setcounter{equation}{0}
\setcounter{figure}{0}
\setcounter{table}{0}
\setcounter{page}{1}
\makeatletter
\renewcommand{\theequation}{S\arabic{equation}}
\renewcommand{\thefigure}{S\arabic{figure}}
\renewcommand{\bibnumfmt}[1]{[S#1]}
\renewcommand{\citenumfont}[1]{S#1}

\begin{figure}[h]
	\includegraphics[]{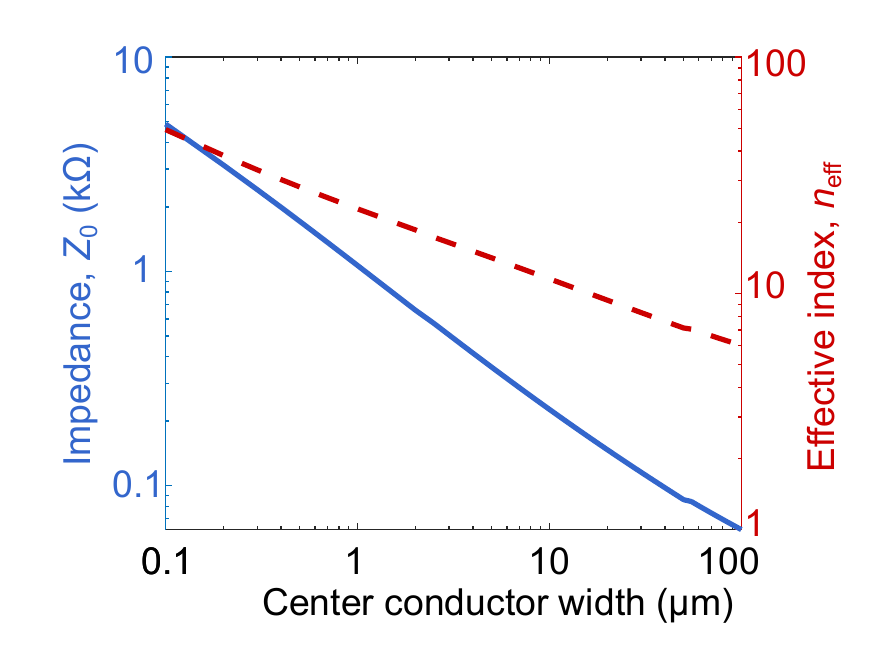}
	\caption{Simulated characteristic impedance ($Z_0$) and effective index ($n_\mathrm{eff}$) of the superconducting NbN coplanar waveguide. The gap size of the coplanar waveguide is fixed at 3 $\upmu$m, and the NbN film has a sheet kinetic inductance of 80\,pH/sq. The substrate is 300\,nm SiO$_2$ on 500\,$\upmu$m intrinsic Si.}
	\label{fig:impedance}
\end{figure}

\begin{figure}[h]
	\includegraphics[]{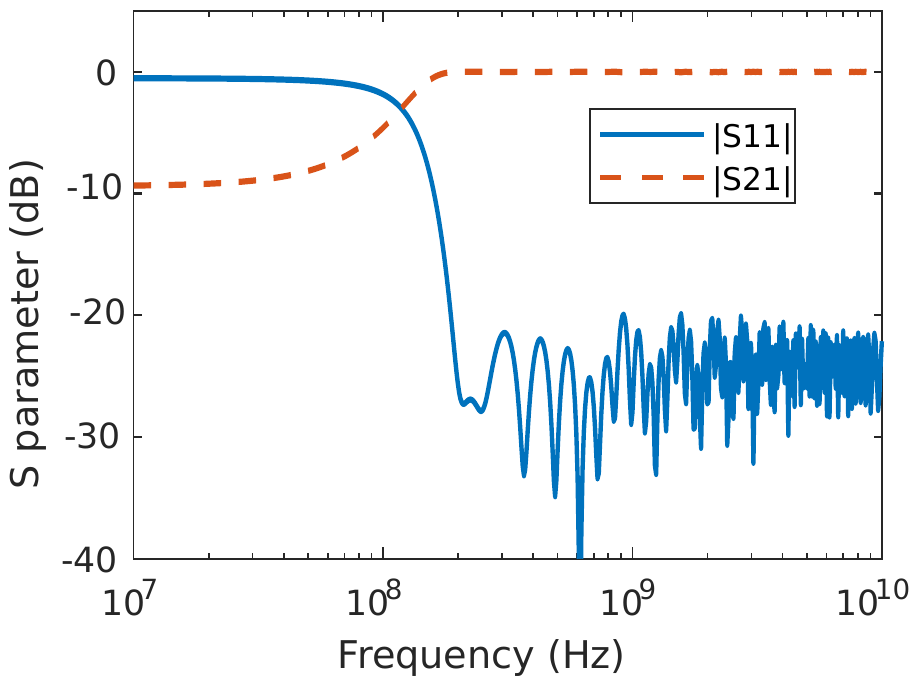}
	\caption{SPICE simulated S-parameter of the impedance taper. In the simulation, the taper is discretized and down-sampled to 300 sections, and each section is implemented using the LTRA model in LTspice. The taper is terminated with impedance matched resistive load on both ends.}
	\label{fig:taper_S_parameter}
\end{figure}

\begin{figure}
	\includegraphics[]{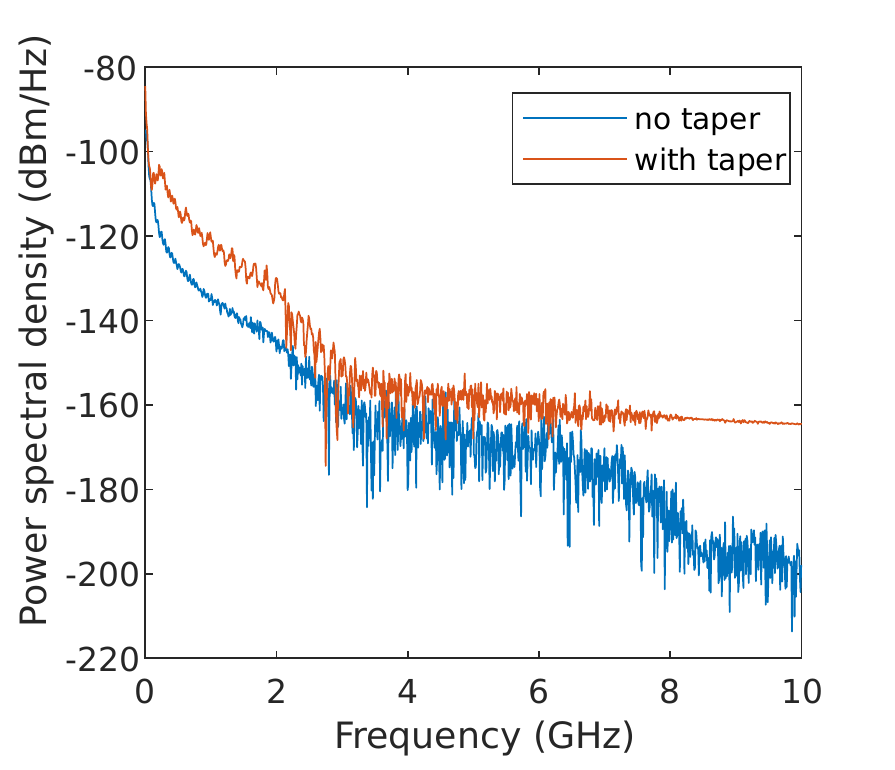}
	\caption{Power spectral density of the output pulses from tapered and reference detectors. The power spectral density was calculated by taking the Fourier transform of the averaged detector pulses acquired on the oscilloscope. The detector pulses were amplified through a 2.5\,GHz amplifier (see main text for the details on measurement setup). The sampling rate was 40 GS/s and the bandwidth was 3\,GHz.}
	\label{fig:psd}
\end{figure}

\begin{figure}
	\includegraphics[width = 3.3in]{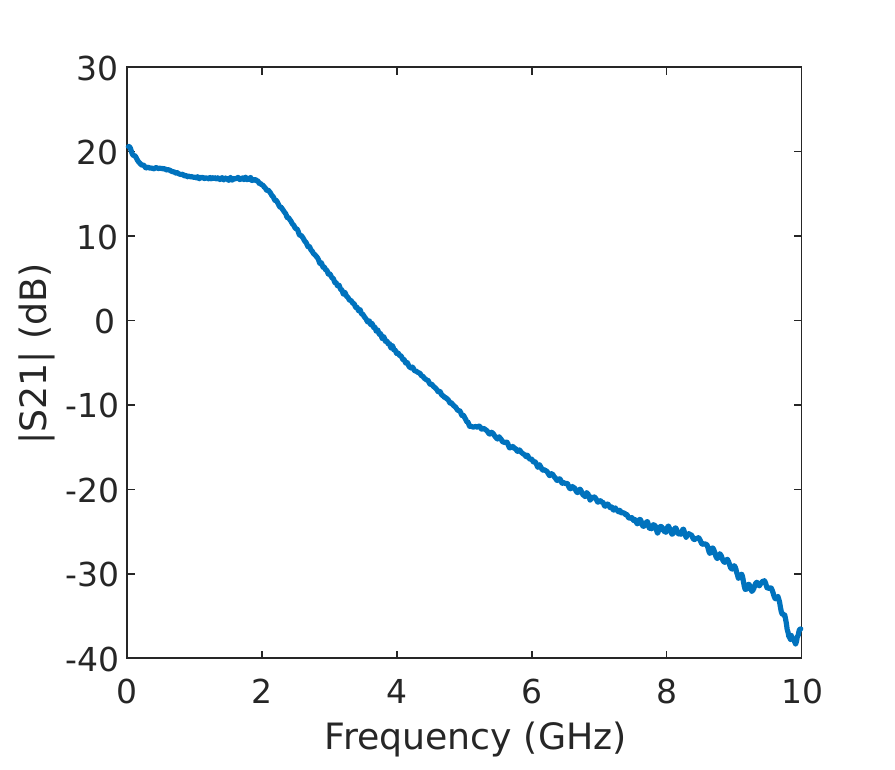}
	\caption{System gain characterization. The transmission coefficient (S21) was measured using a vector network analyzer (Keysight N5224A) from the device under test up to the input port of the oscilloscope, including the cryocable, bias Tee, 3 dB attenuator, and low noise amplifier (LNA2500).}
	\label{fig:gain}
\end{figure}

\begin{figure}
	\includegraphics[width = 6in]{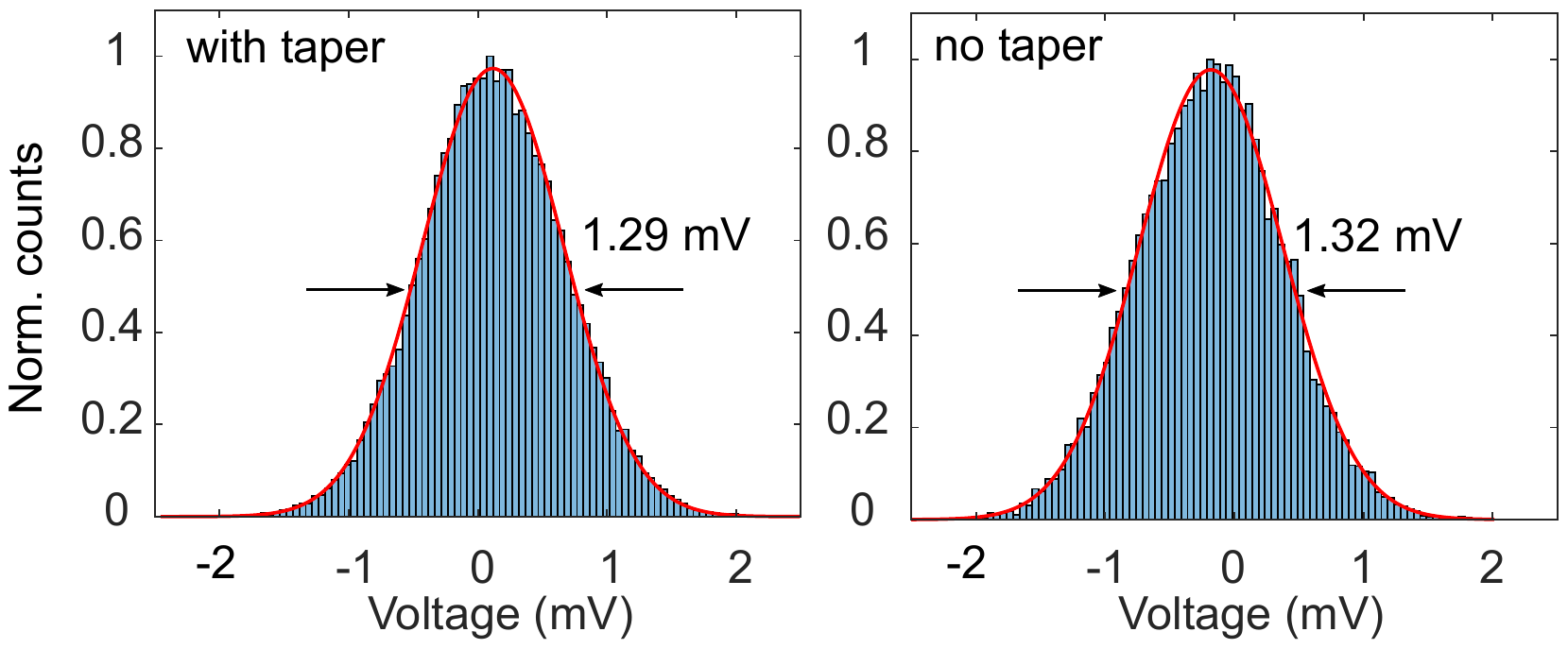}
	\caption{Measured noise floor from the tapered and non-tapered device. Compared to the reference detector, no added noise was observed from the tapered device. The noise voltage was sampled on the oscilloscope at 400\,ps before the rising edge of detector pulses. The sampling rate of the oscilloscope was 40\,GS/s and the bandwidth was set to 3\,GHz.}
	\label{fig:noise_floor}
\end{figure}

\begin{figure}
	\includegraphics[]{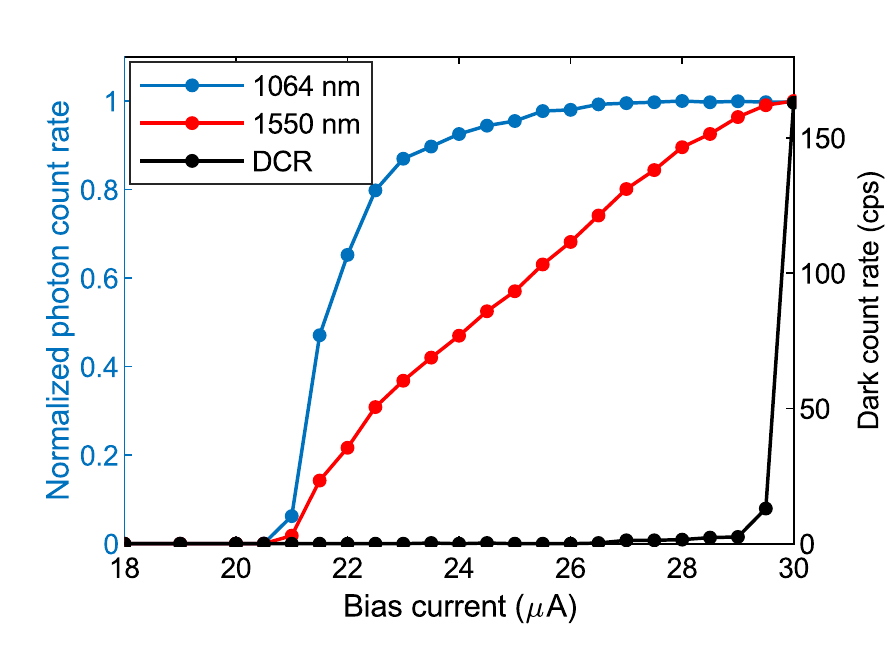}
	\caption{Normalized photon count rate (PCR, left axis) and dark count rate (DCR, right axis) as a function of bias current. At 1064 nm, the detector showed showed saturated internal quantum efficiency. All the pulse shape and jitter measurement reported in work were measured at a bias current of 27.5\,$\upmu$A.}
	\label{fig:PCR}
\end{figure}

\begin{figure}
	\includegraphics[width = 3.3 in]{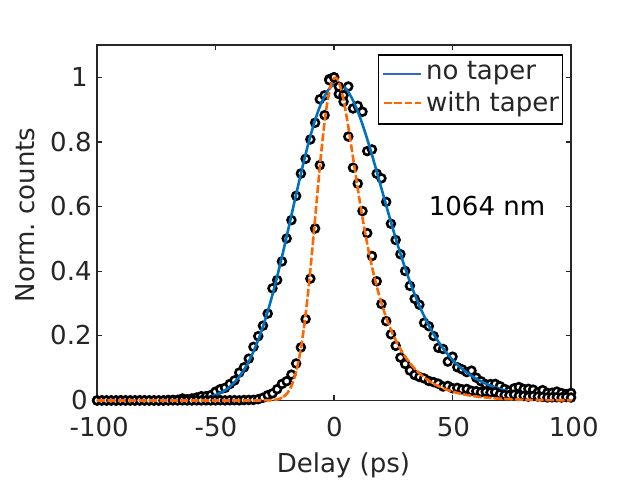}
	\caption{Instrument response function (IRF) under 1064 nm illumination. The bias current was kept at 27.5\,$\upmu$A. For the reference detector, $\sigma$=16.40\,ps, $\lambda$ = 15.9\,ps, FWHM=47.0\,ps. For the tapered detector, $\sigma$ = 6.2\,ps, $1/\lambda$ = 12.5\,ps, FWHM=22.4\,ps. }
	\label{fig:jitter_1064}
\end{figure}

\begin{figure}
	\centering
	\includegraphics[]{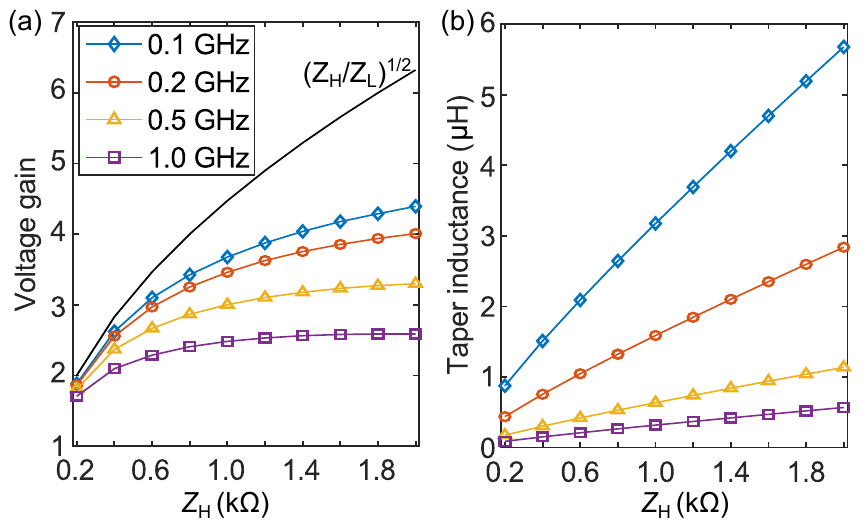}
	\caption{Simulated gain factor (the ratio between the maximum voltages of tapered and non-tapered detector) and taper inductance as a function of the starting input impedance ($Z_\mathrm{H}$) under different nominal cut-off frequencies. Higher input impedance and smaller cut-off frequencies produce higher output voltages (a), but result in larger inductance (b), and hence longer rest time. The non-tapered SNSPD had a bias current of 27.5 $\upmu$A, an inductance of 414 nH, and a maximum output voltage of 1.1 mV (84\%$I_\mathrm{B}\times50\,\Omega$). Each simulated taper had 300 sections and the electrical length was designed to be $l_\mathrm{e}= \mathrm{arccosh}(\rho_0/\rho_\mathrm{pb})c/2\pi f_\mathrm{co}$, where $\rho_0=0.5\ln(Z_\mathrm{H}/Z_\mathrm{L})$, and $\rho_\mathrm{pb}$=0.1, as described in the main text. $Z_\mathrm{L}$ was set to 50\,$\Omega$.} 
	\label{fig:taper_sweep}
\end{figure}

\end{document}